\documentclass[conference]{IEEEtran}
\usepackage[left=0.625in,right=0.625in,top=0.75in,bottom=1in]{geometry}
\usepackage[T1]{fontenc}
\usepackage[latin9]{inputenc}
\usepackage{amsthm}
\usepackage{amsmath}
\usepackage{graphicx} 
\usepackage{color} 
\usepackage{cite}
\usepackage{algpseudocode}
\usepackage{algorithm}
\usepackage{amssymb}
\usepackage{esint}
\usepackage{algpseudocode}
\usepackage{epstopdf}
\usepackage{multirow}
\usepackage{caption}
\usepackage{subcaption}
\usepackage{makecell}
\usepackage{float}
\usepackage{lipsum}
\usepackage{balance}
\usepackage{ctable}
\usepackage{nicefrac}
\newcommand {\Define} {\stackrel {\Delta} {=}  }

\theoremstyle{plain}
\theoremstyle{plain}

\newcommand{\RNum}[1]{\uppercase\expandafter{\romannumeral #1\relax}}

\newcommand{\comment}[1]{}

\IEEEoverridecommandlockouts

\allowdisplaybreaks

\begin{document}

\title{Zak-OTFS and LDPC Codes\vspace{-0.5em}
}

\author{
   \IEEEauthorblockN{Beyza Dabak\IEEEauthorrefmark{1}\textsuperscript{\textsection}, Venkatesh Khammammetti\IEEEauthorrefmark{1}\textsuperscript{\textsection}, Saif Khan Mohammed\IEEEauthorrefmark{2}, and Robert Calderbank\IEEEauthorrefmark{1}}
   \IEEEauthorblockA{\IEEEauthorrefmark{1}Electrical and Computer Engineering Department, Duke University, USA \\ \IEEEauthorrefmark{2}Department of Electrical Engineering, Indian Institute of Technology Delhi, India \\ beyza.dabak@duke.edu, venkatesh.khammammetti@duke.edu, saifkm@ee.iitd.ac.in, and robert.calderbank@duke.edu\vspace{-1em}
   }
}
\maketitle

\begingroup\renewcommand\thefootnote{\textsection}
\footnotetext{These two authors contributed equally to this work.}
\endgroup

\begin{abstract}
Orthogonal Time Frequency Space (OTFS) is a framework for communications and active sensing that processes signals in the delay-Doppler (DD) domain. It is informed by 6G propagation environments, where Doppler spreads measured in kHz make it more and more difficult to estimate channels, and the standard model-dependent approach to wireless communication is starting to break down. We consider Zak-OTFS where inverse Zak transform converts information symbols mounted on DD domain pulses to the time domain for transmission. Zak-OTFS modulation is parameterized by a delay period $\tau_{p}$ and a Doppler period $\nu_{p}$, where the product $\tau_{p}\nu_{p}=1$. When the channel spread is less than the delay period, and the Doppler spread is less than the Doppler period, the Zak-OTFS input-output relation can be predicted from the response to a single pilot symbol. The highly reliable channel estimates concentrate around the pilot location, and we configure low-density parity-check (LDPC) codes that take advantage of this prior information about reliability. It is advantageous to allocate information symbols to more reliable bins in the DD domain. We report simulation results for a Veh-A channel model where it is not possible to resolve all the paths, showing that LDPC coding extends the range of Doppler spreads for which reliable model-free communication is possible. We show that LDPC coding reduces sensitivity to the choice of transmit filter, making bandwidth expansion less necessary. Finally, we compare BER performance of Zak-OTFS to that of a multicarrier approximation (MC-OTFS), showing LDPC coding amplifies the gains previously reported for uncoded transmission.
\end{abstract}

\vspace{-0.3em}
\section{Introduction}
\vspace{-0.2em}
6G represents an opportunity to revisit the fundamentals of wireless communication, as it becomes more and more difficult to estimate channels, and we encounter Doppler spreads measured in kHz. In this paper, we consider the Orthogonal Time Frequency Space (OTFS) framework for wireless communications, where the OTFS carriers are pulses in the delay-Doppler (DD) domain.  

4G and 5G wireless communication networks use Orthogonal Frequency Division Multiplexing (OFDM). Here, a cyclic prefix is used to create shared eigenfunctions of the time shift group, and we have shared eigenfunctions because the time shift group is commutative. Note that pulses in the time domain serve as geometric modes that are moved around by the Linear Time-Invariant (LTI) channel. In this paper, we focus on doubly-spread channels, where there are no shared eigenfunctions because delay shifts and Doppler shifts do not commute. However, pulses in the delay-Doppler domain serve as geometric modes that are moved around by the Linear Time-Variant (LTV) channel. These are the OTFS carrier waveforms, and Section \ref{sec_zak} describes OTFS modulation.

We consider Zak-OTFS where the inverse Zak transform converts information symbols mounted on DD domain pulses to the time domain (TD) for transmission (see \cite{Zakotfs1} and \cite{Zakotfs2} for more details). The Zak transform converts a TD signal to its DD representation, which is parameterized by a delay period $\tau_p$ and a Doppler period $\nu_p$, where $\tau_p\nu_p=1$. The delay-Doppler realization is quasi-periodic with period $\tau_p$ along the delay axis and period $\nu_p$ along the Doppler axis. A DD pulse is spread over $1/B$ sec along the delay domain and $1/T$ Hz along the Doppler domain where $B$ and $T$ are the bandwidth and time-duration of the OTFS frame. Information symbols modulate DD pulses which are located at regular intervals of $1/B$ and~$1/T$ along the delay and Doppler axis respectively, we refer to these DD pulse locations as DD bins. Section \ref{sec_zak} provides more details. 

The Zak-OTFS input-output (I/O) relation is predictable and non-fading when the delay period is greater than the channel delay spread and the Doppler period is greater than the Doppler spread, a condition referred to as the \textit{crystallization condition} in \cite{Zakotfs1, Zakotfs2}. Note that since $\tau_p\nu_p=1$, we are considering underspread channels. We emphasize that it is the interaction between the LTV channel and the OTFS modulation that is predictable and non-fading. Predictability means that reliable communication is possible even when we do not have the fine delay-Doppler resolution necessary to estimate the underlying physical channel. In Section \ref{sec_results} we demonstrate the feasibility of model-free low-density parity-check (LDPC) coded transmission over a Veh-A channel model \cite{Veh_A}, where we do not have enough resources to estimate the channel. 

6G propagation environments are changing the balance between time-frequency methods characteristic of OFDM signal processing and delay-Doppler methods. In OFDM, once the I/O relation is known, equalization is relatively simple, at least when there is no inter-carrier interference (ICI) \cite{ofdm1, OFDMICI}. However, acquisition of the I/O relation is non-trivial and model-dependent. In contrast, equalization is more involved in OTFS, due to intersymbol interference, but acquisition of the I/O relation is simple and model-free (it can be read off from the response to a single pilot waveform \cite{Zakotfs2}). Acquisition becomes more critical in 6G, as Doppler spreads measured in kHz make it more and more challenging to estimate channels. 

\begin{figure*}[htb!]
\centering
\begin{subfigure}{0.30\textwidth}
  \centering
  \includegraphics[width=0.98\linewidth]{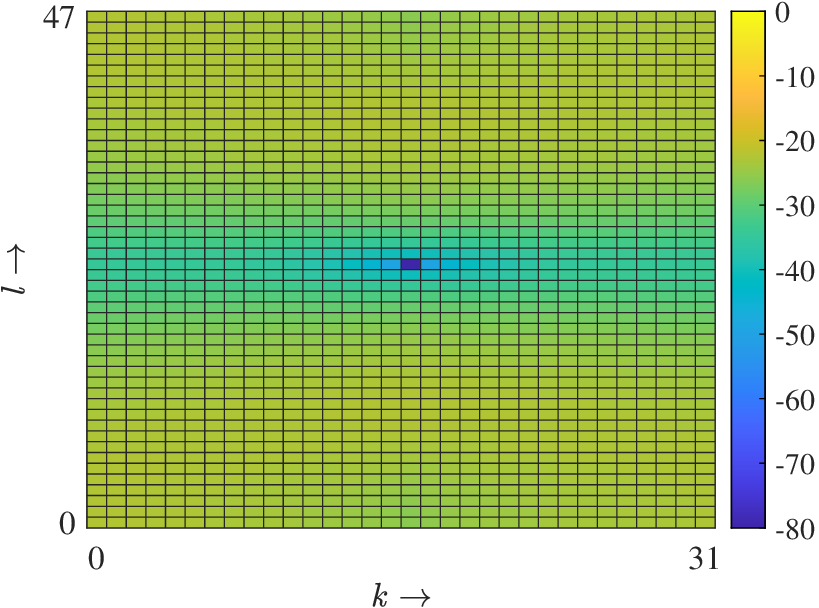}
  \caption{$\nu_{max}=500$Hz}
  \label{rpe_500}
\end{subfigure}%
\hspace{1em}
\begin{subfigure}{0.30\textwidth}
  \centering
  \includegraphics[width=0.98\linewidth]{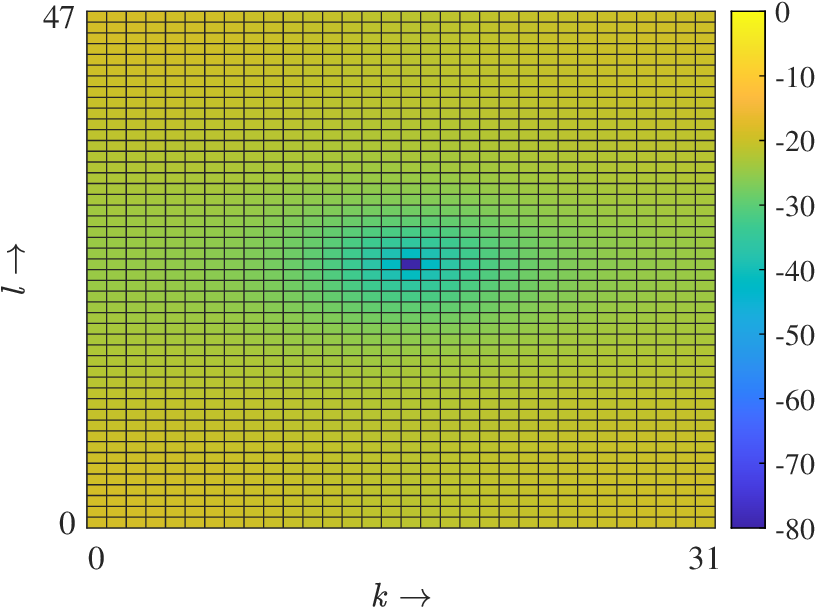}
  \caption{$\nu_{max}=4500$Hz}
  \label{rpe_4500}
\end{subfigure}%
\hspace{1em}
\begin{subfigure}{0.30\textwidth}
  \centering
  \includegraphics[width=0.98\linewidth]{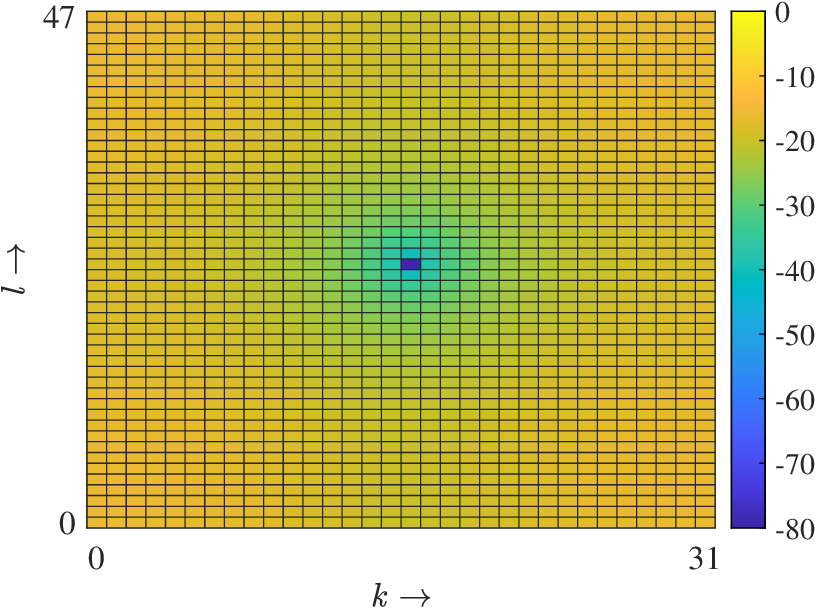}
  \caption{$\nu_{max}=12000$Hz}
  \label{rpe_12000}
\end{subfigure}
\vspace{-0.4em}
\caption{Heatmap of RPE in the crystallization regime for $\nu_{p}=30$kHz, $\tau_{p} = 1/\nu_{p}$, $B=0.96$MHz, $T=1.6$ms, Veh-A Channel for varying maximum Doppler shifts $\nu_{max}$.}
\label{nup30K_vehA}
\vspace{-0.95em}
\end{figure*}

We follow \cite{Zakotfs1, Zakotfs2} in predicting the effective channel response from the response to a single pilot signal. When Doppler spreads increase, the number of reliable DD bins for which the received response to a symbol transmitted on that bin can be reliably estimated from the response to a single pilot, decreases, and the reliable DD bins concentrate around the pilot location. This is illustrated in Fig. \ref{nup30K_vehA} with heatmaps of the relative prediction error (RPE) for the channel response to a transmitted DD signal. The RPE is the error in estimating the effective DD domain channel response to an arbitrary transmit signal based on the DD domain response to a single pilot signal. Section III provides more details about RPE. We use this prior information about the geometry of the reliable DD bins to configure LDPC  codes. 

LDPC codes are specified by graphs and are the error correction technique of choice in many communications and data storage contexts\cite{modern_coding_theory}. Message-passing decoders diffuse information carried by parity bits into the payload, and in \cite{udp}, the authors show that LDPC decoders benefit from the differences between reliabilities of parity and message bits using density evolution analysis. In data storage applications there is a benefit to engineering parity bits to be more reliable than message bits \cite{udp}. In other applications, where the reliabilities are less commensurable, there is a benefit to mapping information bits of LDPC (and Turbo) codes to higher reliability QAM constellations \cite{uep_ldpc, uep_turbo}. We now highlight our main contributions.

\textbf{Resilience to Doppler Spread:} When the crystallization conditions hold, we show that LDPC coding extends the range of Doppler spreads for which reliable model-free communication is possible on a Veh-A channel model.

\textbf{LDPC Code Configuration:} We show that it is advantageous to allocate information symbols to the more reliable DD bins in the DD domain. 

\textbf{Transmit and Receive Filters:} Prior work on uncoded OTFS transmission \cite{Zakotfs1, Zakotfs2} had observed that reliability was significantly improved by substituting a root raised cosine (RRC) filter for a sinc filter, at the cost of higher OTFS frame duration and bandwidth. We show that LDPC coding reduces the difference between RRC and sinc filters, making bandwidth expansion less necessary. 

\textbf{Optimality of Zak-OTFS:} Over the past few years, several variants of OTFS have been reported in the literature \cite{TharajViterbo,Bestreads}. A multicarrier approximation to Zak-OTFS, which we refer to as MC-OTFS has been the focus of most research attention so far \cite{OTFSOFDM,Hadaniwhitepaper}. Prior work on uncoded transmission \cite{Zakotfs1,Zakotfs2} had observed that the MC-OTFS I/O relation was less predictable than the Zak-OTFS I/O relation, and as a consequence bit error rate (BER) performance was inferior. We show that LDPC coding amplifies these gains.

\vspace{-0.5mm}
\section{Zak-OTFS Modulation} \label{sec_zak}
\vspace{-0.5mm}
\begin{figure*}[h]
    \centering
    \includegraphics[clip=true, trim=0in 0in 0in 0.4in, width=0.65\textwidth]{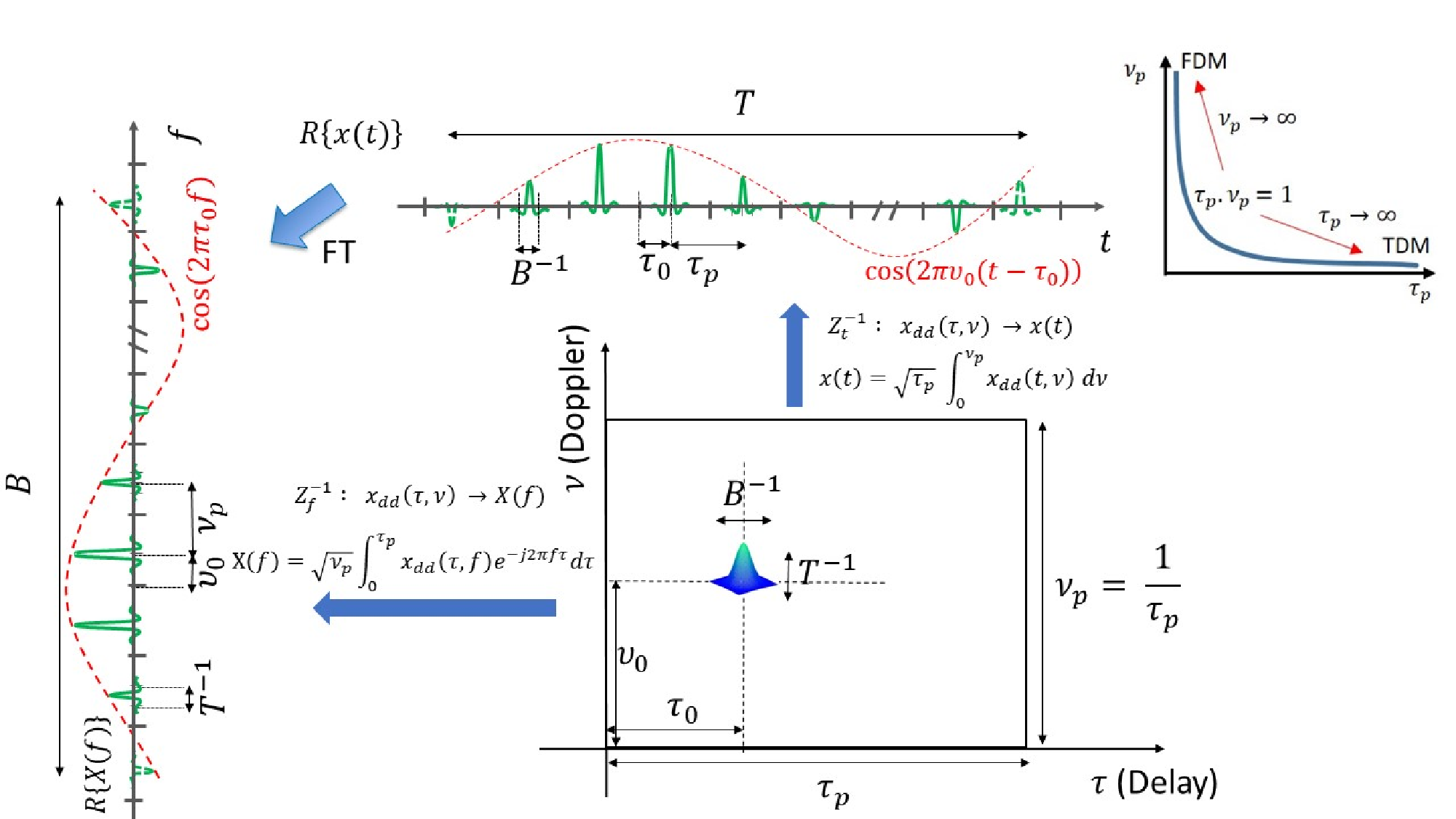}
    \caption{A DD domain pulse and its time domain/frequency domain realizations referred to as TD/FD pulsone. The TD pulsone comprises of a finite duration pulse train modulated by a TD tone. The FD pulsone comprises of a finite bandwidth pulse train modulated by a FD tone. The location of the pulses in the TD/FD pulse train and the frequency of the modulated TD/FD tone is determined by the location of the DD domain pulse $(\tau_0, \nu_0)$. As the Doppler period $\nu_p\to\infty$, the FD pulsone approaches a single FD pulse which is the FDM carrier. Similarly, as the delay period $\tau_p\to\infty$, the TD pulsone approaches a single TD pulse which is the TDM carrier. Setting $\tau_p\nu_p = 1$, we see that OTFS is a family of modulations parameterized by $\tau_p$ that interpolates between TDM and FDM \cite{Zakotfs1}.}
    \label{fig4paper1} 
    \vspace{-1em}
\end{figure*}

A pulse in the DD domain is a quasi-periodic localized function, defined by a delay period $\tau_p$ and a Doppler period $\nu_p$.
In the \emph{period lattice} $\Lambda_p = \{ (n \tau_p, m \nu_p) \, | \, n,m \in {\mathbb Z} \}$, there is only one pulse within the \emph{fundamental region} ${\mathcal D}_0 = \{ (\tau, \nu) \, | \, 0 \leq \tau < \tau_p, 0 \leq \nu < \nu_p \},$ and there are infinitely many replicas along the delay and Doppler axes given by \vspace{-1mm}
\begin{eqnarray}
    x_{_{\mbox{\scriptsize{dd}}}}(\tau + n \tau_p,\nu + m \nu_p)  & = &   e^{j 2 \pi n \nu \tau_p} \, x_{_{\mbox{\scriptsize{dd}}}}(\tau,\nu),
\end{eqnarray}\vspace{-5.5mm} \\ for all $n,m \in {\mathbb Z}$. Only
quasi-periodic DD domain functions can have a TD representation. When viewed in the time domain, this function is realized as a pulse train modulated by a tone (see Fig. \ref{fig4paper1} taken from \cite{Zakotfs1}), hence the name \textit{pulsone}. The DD domain pulse is the orthogonal time frequency space (Zak-OTFS) waveform, introduced in~\cite{OTFSOFDM}, and widely studied thereafter.

\vspace{-0.4em}
\subsection*{Zak-OTFS modulation:} 
\vspace{-0.1em}
We obtain a DD domain pulse by applying a factorizable pulse shaping waveform $w_{tx}(\tau,\nu)=w_{1}(\tau)w_{2}(\nu)$ to a quasi-periodic Dirac-delta DD domain pulse. The spread of $w_{1}(\tau)$ along the delay axis is approximately $1/B$ where $B$ is the bandwidth of the TD carrier waveform. Similarly, the spread of $w_{2}(\nu)$ along the Doppler axis is approximately $1/T$ where $T$  is the time duration of the TD carrier. Fig. \ref{fig4paper1} illustrates how the inverse Zak transform converts the pulse in the DD domain to a TD waveform. 
Since each DD pulse repeats quasi-periodically, there are $M = \tau_p / (1/B) = B \tau_p$ pulse locations along the delay axis and $N = \nu_p / (1/T) = T \nu_p$ pulse locations along the Doppler axis and the $MN$ pulses at these pulse locations
have minimal overlap, rendering OTFS an orthogonal modulation that achieves the Nyquist rate. The pulse locations are the points of the \emph{information lattice}  \vspace{-2mm}
\begin{eqnarray}
    \Lambda_{_{\mbox{\scriptsize{dd}}}}  & \Define &  \{ (k/B, l/T) \, | \, k,l \in {\mathbb Z} \}.
\end{eqnarray} \vspace{-5mm} \\
The $MN$ quasi-periodic pulses located at information lattice points within ${\mathcal D}_0$ carry the $MN$ information symbols $x[k,l]$, $k=0,1,...,M-1$, $l=0,1,...,N-1$. Here $x[k,l]$ is carried by the pulse at $(k/B,l/T)$. \\
\vspace{-0.8em}
\subsubsection*{Zak-OTFS transceiver signal processing} 
Let \vspace{-1mm}
\begin{equation}
x_{_{\mbox{\scriptsize{dd}}}}[k+n M, l+m N]  \Define  x[k, l] e^{j 2 \pi n \frac{l}{N}}, \textup{ } m, n \in \mathbb{Z},
\end{equation}
\begin{eqnarray}
\label{eqnxddkl}
    x_{_{\mbox{\scriptsize{dd}}}}(\tau,\nu) & \Define & \sum\limits_{k,l \in {\mathbb Z}} x_{_{\mbox{\scriptsize{dd}}}}[k,l] \, \delta(\tau - k/B) \, \delta(\nu - l/T),
\end{eqnarray}
$x_{_{\mbox{\scriptsize{dd}}}}(\tau,\nu)$ denote the quasi-periodic information signal which is a superposition of all $MN$ quasi-periodic Dirac-delta pulses, wherein each such quasi-periodic pulse is modulated by an information symbol and is located at one of the $MN$ distinct points of the information lattice within the fundamental region ${\mathcal D}_0$. The filtered information signal is given by \vspace{-1.3mm}
\begin{eqnarray}
\label{eqn156}
x_{_{\mbox{\scriptsize{dd}}}}^{w_{tx}}(\tau,\nu) & = & w_{tx}(\tau, \nu) \, *_{\sigma} \, x_{_{\mbox{\scriptsize{dd}}}}(\tau,\nu),
\end{eqnarray} \vspace{-5.4mm} \\ 
where `$*_{\sigma}$' represents \emph{twisted convolution}\footnote{$
a(\tau, \nu) *_\sigma b(\tau, \nu) \triangleq \iint a\left(\tau^{\prime}, \nu^{\prime}\right) b\left(\tau-\tau^{\prime}, \nu-\nu^{\prime}\right) e^{j 2 \pi \nu^{\prime}\left(\tau-\tau^{\prime}\right)} d \tau^{\prime} d \nu^{\prime} $}.

At the receiver, the DD domain representation $y_{_{\mbox{\scriptsize{dd}}}}(\tau,\nu)$ of the received TD signal $y(t)$ is matched-filtered with a receive DD domain filter $w_{rx}(\tau, \nu)$. The filtered received DD domain signal is given by \vspace{-1.1mm}
\begin{eqnarray}
\label{eqn157} 
y_{_{\mbox{\footnotesize{dd}}}}^{w_{rx}}(\tau, \nu) & = & w_{rx}(\tau, \nu) \, *_{\sigma} \, y_{_{\mbox{\footnotesize{dd}}}}(\tau, \nu) + n(\tau,\nu),
\end{eqnarray} \vspace{-4.5mm} \\ 
where $n(\tau,\nu)$ is the DD additive white Gaussian noise (AWGN) at the receiver with power spectral density (PSD) (of its TD representation) $N_0$. It is shown in \cite{Zakotfs1} that the doubly-spread channel acts through twisted convolution, i.e., \vspace{-1mm}
\begin{eqnarray}
\label{eqn158}
y_{_{\mbox{\footnotesize{dd}}}}(\tau, \nu) & = & h_{_{\mbox{\scriptsize{phy}}}}(\tau, \nu) \, *_{\sigma} \, x_{_{\mbox{\scriptsize{dd}}}}^{w_{tx}}(\tau,\nu)+ n(\tau,\nu).
\end{eqnarray}\vspace{-4.5mm} \\  
The sparse delay-Doppler channel spreading function $h_{_{\mbox{\scriptsize{phy}}}}(\tau, \nu)$ is given by \vspace{-4mm}
\begin{eqnarray}
    h_{_{\mbox{\scriptsize{phy}}}}(\tau, \nu)=\sum_{i=1}^{p} h_i \delta\left(\tau-\tau_i\right) \delta\left(\nu-\nu_i\right), 
\end{eqnarray} \vspace{-3mm} \\ 
where $\delta(\cdot)$ denotes the Dirac-delta impulse function and $p$ is the number of paths. 
From (\ref{eqn156}), (\ref{eqn157}) and (\ref{eqn158}) it follows that
\vspace{-1.4mm}
\begin{eqnarray}
\label{eqn159}
y_{_{\mbox{\footnotesize{dd}}}}^{w_{rx}}(\tau, \nu) \hspace{-3mm} & = & \hspace{-3mm}{\Bigg (} w_{rx}(\tau, \nu) \, *_{\sigma} \, {\Big [}
h_{_{\mbox{\scriptsize{phy}}}}(\tau, \nu) \, *_{\sigma} \, \nonumber  \\ 
& &  {\Big (} w_{tx}(\tau, \nu) \, *_{\sigma} \, x_{_{\mbox{\scriptsize{dd}}}}(\tau,\nu) {\Big )} \, {\Big ]} \, {\Bigg )}+ n(\tau,\nu).
\end{eqnarray} 
Since twisted convolution is associative 
\begin{eqnarray}
\label{eqn160}
y_{_{\mbox{\footnotesize{dd}}}}^{w_{rx}}(\tau, \nu) \hspace{-3mm} & = & \hspace{-3mm} \, h_{_{\mbox{\scriptsize{eff}}}}(\tau, \nu)*_{\sigma} \, x_{_{\mbox{\scriptsize{dd}}}}(\tau,\nu)+ n(\tau,\nu), \\
h_{_{\mbox{\scriptsize{eff}}}}(\tau, \nu) \hspace{-3mm} & \Define & \hspace{-3mm} {\Big (}  w_{rx}(\tau, \nu) \, *_{\sigma} \,  h_{_{\mbox{\scriptsize{phy}}}}(\tau, \nu) \, *_{\sigma} \, w_{tx}(\tau, \nu) {\Big )}.
\end{eqnarray}
At the receiver, the quasi-periodic filtered signal $y_{_{\mbox{\footnotesize{dd}}}}^{w_{rx}}(\tau, \nu)$ is sampled on the information lattice resulting in the discrete DD domain received signal \vspace{-2mm}
\begin{eqnarray}
\label{ch2_eqn5paper}
y_{_{\mbox{\footnotesize{dd}}}}[k,l] &  = &   y_{_{\mbox{\footnotesize{dd}}}}^{w_{rx}}\left(\tau = k \frac{\tau_p}{M}, \nu = l \frac{\nu_p}{N} \right)
\end{eqnarray}
for all $k,l \in {\mathbb Z}$. A quasi-periodic DD domain signal when sampled on the information lattice results in a discrete DD domain quasi-periodic signal that satisfies the discrete quasi-periodic property, \vspace{-1mm}
\begin{eqnarray}
    y_{_{\mbox{\footnotesize{dd}}}}[k +nM,l + mN] & = & e^{j 2 \pi \frac{ n l}{N}} \, y_{_{\mbox{\footnotesize{dd}}}}[k,l],
\end{eqnarray}
for all $k,l,n,m \in {\mathbb Z}$.

From (\ref{eqnxddkl}), (\ref{eqn160}) and (\ref{ch2_eqn5paper}), the Zak-OTFS I/O relation between the discrete DD domain input signal $x_{_{\mbox{\footnotesize{dd}}}}[k,l], k,l \in {\mathbb Z}$ and the discrete output
signal $y_{_{\mbox{\footnotesize{dd}}}}[k,l]$ is given by \cite{Zakotfs2} \vspace{-1mm}
\begin{eqnarray}{\label{rx_zak_DD_signal}}
y_{_{\mbox{\footnotesize{dd}}}}[k,l] \hspace{-1mm} &=& \hspace{-1mm} \sum_{k^{\prime}, l^{\prime} \in \mathbb{Z}} h_{\mathrm{eff}}\left[k-k^{\prime}, l-l^{\prime}\right] x_{\mathrm{dd}}\left[k^{\prime}, l^{\prime}\right] e^{j 2 \pi \frac{\left(l-l^{\prime}\right)}{N} \frac{k^{\prime}}{M}} \nonumber \\
& & \hspace{52mm} + n[k,l] \nonumber \\
& = & \left(h_{_{\mbox{\scriptsize{eff}}}}[k,l] \, *_{\sigma} \, x_{_{\mbox{\footnotesize{dd}}}}[k,l]\right) + n[k,l], \nonumber \\
h_{_{\mbox{\scriptsize{eff}}}}[k,l] & \Define & h_{_{\mbox{\scriptsize{eff}}}}\left(  \tau = \frac{ k \tau_p}{M} \,,\, \nu = \frac{l \nu_p}{N} \right).
\end{eqnarray}
This is simply the discrete twisted convolution between the effective discrete DD domain channel filter $h_{_{\mbox{\scriptsize{eff}}}}[k,l]$ and the discrete input signal $x_{_{\mbox{\scriptsize{dd}}}}[k,l]$.

\subsection*{Optimality of Zak-OTFS:} 
\vspace{-0.1em}
Over the past few years, several variants of OTFS have been reported in the literature \cite{TharajViterbo}. A multicarrier approximation to Zak-OTFS, which we refer to as MC-OTFS, has been the focus of most research attention so far \cite{Bestreads, OTFSOFDM, Hadaniwhitepaper}. We have seen that Zak-OTFS uses the inverse Zak transform to convert a DD domain pulse to a TD carrier waveform in a single step. By contrast, MC-OTFS converts a DD domain pulse to a TD carrier waveform in two steps. The first step is to use the inverse symplectic finite Fourier transform (ISFFT) to convert from the DD domain to the TF domain, and the second step is to use the Heisenberg transform to convert from the TF domain to TD.

We have seen that the I/O relation for Zak-OTFS is a simple twisted convolution of effective channel and input. By contrast, the I/O relation for MC-OTFS is a much more complicated mixture of periodic and quasi-periodic signal processing (see \cite{Zakotfs2} for details). As a consequence, the I/O relation for MC-OTFS is less predictable than that of Zak-OTFS. As the Doppler spread increases, the BER performance of MC-OTFS is inferior to that of Zak-OTFS \cite{Zakotfs2}. When the crystallization conditions hold, model-free operation with Zak-OTFS is successful, and performance is only slightly worse than performance with perfect knowledge of the effective channel. By contrast, model-free operation with MC-OTFS is not feasible \cite{Zakotfs2}.
\vspace{-0.1em}
\section{Configuring LDPC Codes} \label{sec_model}
\vspace{-0.1em}

Prior work on multiple access with MC-OTFS \cite{OMA_otfs} explored how to allocate information symbols in the DD domain to minimize interference in the time-frequency domain. Although that is a different problem, it is possible to design allocation strategies to improve spectral efficiency and BER performance \cite{OMA_otfs, NOMA_otfs}.
\vspace{-0.1em}
\subsection{Relative Prediction Error (RPE)}\label{subsec_rpe}
\vspace{-0.1em}
RPE is the error in estimating a complex received signal from the pilot response.  The precise mathematical definition is given in \cite{Zakotfs2}, Equation (33). We estimate the channel response based on a single pilot transmitted on a DD pulse located at the center $(M/2, N/2)$ of the fundamental DD period, as in\cite{Zakotfs2}. We consider a 6-path Veh-A channel model\cite{Veh_A} with OTFS signal bandwidth $B = 0.96$MHz and time duration $T = 1.6$ ms. Power-delay profile of the six channel paths is shown in Table \ref{table1}, and we note that the paths are not separable (the difference in their delays is less than the delay domain resolution $1/B$) and cannot be estimated accurately. The crystallization conditions hold ($\nu_{p}=30$kHz), we assume sinc transmit and receive filters $w_{tx}(\tau, \nu) = w_{rx}(\tau, \nu) = \sqrt{B T} sinc(B \tau) sinc(T \nu)$, and we take $M=32$, $N=48$. Figure \ref{nup30K_vehA} provides three heatmaps that illustrate how RPE changes as we vary the Doppler spread.

\begin{table}
\caption{Power-Delay Profile of Veh-A Channel 
\cite{Veh_A}.}
\vspace{-0.4em}
\centering
\begin{tabular}{|c!{\vrule width 0.1em}c|c|c|c|c|c|}
\hline
Path number $i$ & $1$ & $2$ & $3$ & $4$ & $5$ & $6$   \\
\specialrule{.1em}{.05em}{.05em} 
Rel. Delay $\tau_{i}$ $(\mu s)$ & $0$ & $0.31$ & $0.71$ & $1.09$ & $1.73$ & $2.51$   \\
\hline
Rel. Power $\frac{{\mathbb E}[ | h_{i} |^2 ]}{{\mathbb E}[ | h_{1} |^2 ]}$ (dB) & $0$ & $-1$ & $-9$ & $-10$ & $-15$ & $-20$    \\
\hline
\end{tabular}
\label{table1}
\vspace{-1.5em}
\end{table}

When the Doppler spread is low (Fig. \ref{rpe_500}), we observe a rectangle of highly reliable symbols\footnote{Symbols correspond to QAM symbols.}, centered at the pilot location, extending over the entire delay domain.  At intermediate Doppler spreads (Fig. \ref{rpe_4500}), symbols at the edge of this rectangle become less reliable, and at high Doppler spreads (Fig. \ref{rpe_12000}), highly reliable symbols are concentrated around the pilot symbol. The precise symbol reliabilities will vary as we consider different channels, but we assume that the region of highly reliable symbols will evolve in a broadly similar way as we vary the Doppler spread. It is this pattern of evolution that informs our allocation strategy.
\vspace{-0.30em}
\subsection{Allocation Strategy} {\label{Proposed_alloc}}
\vspace{-0.25em}

The observations made from the RPE heatmaps offer an opportunity to extend the range of Doppler spreads for reliable operation via LDPC code configuration by exploiting the differences in reliabilities. As mentioned before, \cite{udp} shows that LDPC decoders can take advantage of the differences between reliabilities of parity and message bits. Thus, \textit{we propose allocation schemes where we place the transmit information symbols around the pilot (more reliable locations) and parity symbols away from the pilot (less reliable locations)}.

We consider two allocation strategies, which we refer to as \textit{RPE allocation} and \textit{strip allocation}. In RPE allocation we list the RPE values in ascending order, so the most reliable symbol is listed first. Given an LDPC code with $k$ information symbols, we allocate them to the $k$ most reliable DD bins. Thus, information symbols correspond to lower RPE values and parity symbols correspond to higher RPE values. This allocation method is of theoretical interest, but we will not be able to access the exact RPE values in a wireless system. Thus, we propose strip allocation based on the observations made from Fig. \ref{nup30K_vehA}. In strip allocation, we allocate information symbols to a rectangle, centered at the pilot location, that extends over the entire delay domain.  We choose the height of the rectangle to encompass all the information symbols. The remaining DD bins (outside the rectangle) are allocated to parity symbols of the LDPC codeword. Strip allocation is oblivious to the precise RPE values, and simply looks to take advantage of the pattern of evolution (see Fig. \ref{nup30K_vehA}). Our simulation results suggest that the performance of strip and RPE allocations are very close.
\vspace{-0.25em}
\subsection{Vectorized I/O Relation} 
\vspace{-0.25em}
In the standard allocation scheme, the codeword bits generated from the LDPC encoder, for a given rate, are converted to QAM symbols $x[k,l]$.  These symbols are then allocated the DD resources of the 2-D grid $x_{_{\mbox{\footnotesize{dd}}}}[k,l]$ (for \,\ $k=0,1,\cdots,M-1, \,\ l=0,1,\cdots,N-1$). But in the proposed allocation scheme, we allocate the more reliable DD bins to the information symbols and then allocate the remaining DD bins to the parity symbols. By using Zak-OTFS modulation, we transform these symbols into TD and transmit it over the air. The receiver receives the TD signal and demodulates to $y_{_{\mbox{\scriptsize{dd}}}}[k,l]$. The vector notation of  $y_{_{\mbox{\scriptsize{dd}}}}[k,l]$ in (\ref{rx_zak_DD_signal}) is given by
\begin{eqnarray}{\label{vecor_notate}}
\mathbf{y}_{_{\mbox{\footnotesize{dd}}}} &=& \mathbf{H}_{_{\mbox{\footnotesize{dd}}}}\mathbf{x}_{_{\mbox{\footnotesize{dd}}}}+\mathbf{n}_{_{\mbox{\footnotesize{dd}}}},
\end{eqnarray}
where $\mathbf{y}_{_{\mbox{\footnotesize{dd}}}} \in \mathbb{C}^{MN \times 1}$ and the $(lM + k + 1)$-th element of $\mathbf{y}_{_{\mbox{\footnotesize{dd}}}}$ is $y_{_{\mbox{\scriptsize{dd}}}}[k,l]$. Also, $\mathbf{H}_{_{\mbox{\footnotesize{dd}}}} \in \mathbb{C}^{MN \times MN}$ where $h_{\mathrm{eff}}\left[k-k^{\prime}, l-l^{\prime}\right]$ is the element in $(l^{\prime}M + k^{\prime} + 1)$-th row and $(lM + k + 1)$-th column  of $\mathbf{H}_{_{\mbox{\footnotesize{dd}}}}$.

\section{Results} \label{sec_results}
\vspace{-0.2em}
\subsection{Simulation Framework}
\vspace{-0.3em}
We present results for the 6-path Veh-A channel model~introduced in Section \ref{subsec_rpe}. The delay-power profile of the channel is shown in Table \ref{table1}. Complex channel gains $h_{i}$ are independent Rayleigh distributed random variables, and we normalize these gains by setting $\sum\nolimits_{i=1}^{6} {\mathbb E}[| h_{i} |^2] = 1 $.  For $i = 1,2,\cdots,6$, we model the Doppler shift associated with the $i$-th path as $\nu_{i} = \nu_{max}\cos\left(\theta_{i}\right)$ where $\theta_{i}$ is independent and uniformly distributed in $[0, 2\pi)$. For the simulations, we ensured that the crystalline condition for Zak-OTFS modulation was satisfied.  

We transmit a single pilot at the center of the fundamental DD period ${\mathcal D}_0$, and we estimate the channel matrix $\widehat{\mathbf{H}}_{_{\mbox{\footnotesize{dd}}}}$ from the received signal (as in \cite{Zakotfs2}). Using the estimated channel, we perform Minimum Mean Squared Error (MMSE) equalization on the received vector $\mathbf{y}_{_{\mbox{\footnotesize{dd}}}}$. The average SNR of the system is $E_{T}/N_{0}$ where $E_{T} = \mathbb{E}\{|\mathbf{x}_{_{\mbox{\footnotesize{dd}}}}[k,l]|^{2}\}$ is the expected energy of an information symbol, and $N_{0}$ is the noise power spectral density of AWGN. We simulate a rate $1/2$ LDPC code of block length $3012$. We started from a binary protograph-based quasi-cyclic code with circulant size $2$, variable node degree $3$, check node degree $6$ with lifting parameter $251$ and decoded on its systematic form with a layered belief propagation (BP) decoder\cite{layered_bp}, and we plot BER curves for 4-QAM modulation.

\vspace{-0.3em}
\subsection{Coded Transmission}
\begin{figure}[htb!]
     \vspace{-1em}
        \centering
        \includegraphics[width=0.9\linewidth]{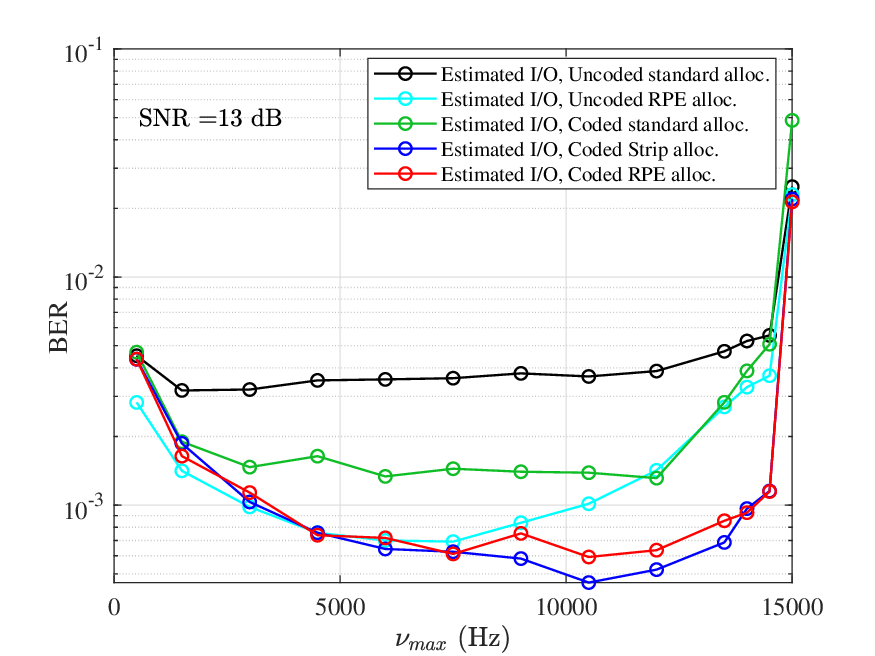}
        \vspace{-0.65em}
        \caption{The effect of coding with proposed allocation (light blue vs. red and dark blue curves), the effect of coding with standard allocation (black vs. green curves), the effect of RPE allocation on uncoded transmission (black vs. light blue curves), and the effect of proposed allocations on coded transmission (green vs. red and dark blue curves).} 
        \label{uncodevscode}
    \end{figure}

\vspace{-0.3em}
We observe the effect of coding both with proposed allocations and without allocation, which we refer to as standard allocation. For the first, our baseline is uncoded transmission at the same rate with the estimated channel and with RPE allocation of information. Note that half the DD domain symbols are unused. We set SNR = $13$ dB, and plot BER as a function of $\nu_{max}$ in Fig. \ref{uncodevscode}. We observe that the BER performance of uncoded transmission with RPE allocation (light blue curve) degrades when $\nu_{max}$
exceeds $10.5$kHz. By contrast, the BER performance of proposed coded transmission (red and dark blue curves) only starts to degrade at about $14$kHz, which is close to the edge of the crystallization regime $(2 \nu_{max} < \nu_{p})$. We can also observe the effect of coding without proposed allocation. Now the baseline is uncoded transmission at the same rate without RPE allocation of information symbols. Note that for uncoded transmission, standard allocation is the same as strip allocation since we allocate all the information symbols (without any message or parity distinction) to a rectangle, centered at the pilot location\footnote{In uncoded RPE allocation, information symbols are placed in DD bins where RPE values are lower. For uncoded transmission, allocation is meaningful when some of the DD domain symbols are unused in the frame.}. In Fig. \ref{uncodevscode}, we observe that coding improves BER for the standard allocation scheme as well (black vs. green curves).

We also observe the effect of allocation for uncoded and coded transmission. It is better to use more reliable symbols to transmit information, and Fig. \ref{uncodevscode} quantifies the gains. For uncoded transmission, in Fig. \ref{uncodevscode} we observe that for a model-free operation, and when the uncoded symbols occupy only half of the total resources (DD bins), the RPE allocation scheme outperforms the standard allocation scheme (black vs. light blue curves). Similarly, for coded transmission, RPE and strip allocation improve BER performance (green vs. red and dark blue curves). We also observe very little difference between the BER performance of strip and RPE allocation. A desired BER of $10^{-3}$ can only be achieved with RPE allocation in uncoded and with proposed allocation strategies in coded transmission (see Fig. \ref{uncodevscode}). Also, the same desired BER of $10^{-3}$ can be achieved even at higher $\nu_{max}$ of around $14500$Hz in proposed allocation schemes for coded transmission i.e., \textit{coding extends the range of Doppler spreads for reliable communication.}

\begin{figure}[htb!]
    \vspace{-2.5em}
        \centering
        \includegraphics[width=0.9\linewidth]{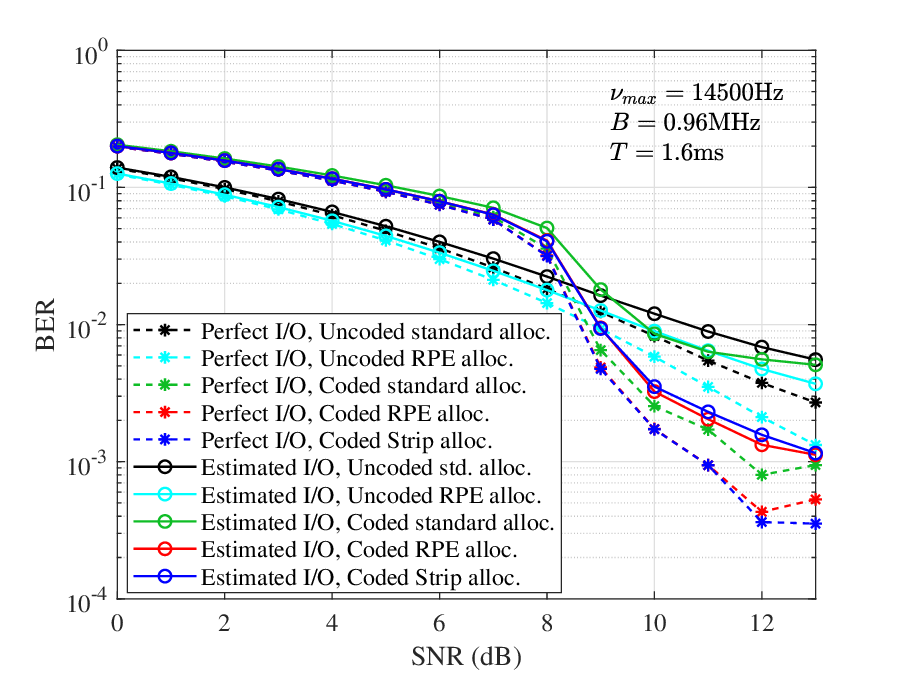}
        \vspace{-0.7em}
        \caption{Comparison between all strategies for varying SNR at $\nu_{max}=14500$Hz. }
        \label{bervssnr}
        \vspace{-1em}
    \end{figure}

Fig. \ref{bervssnr} plots BER as a function of SNR for a $\nu_{max}$ of $14500$Hz. In Fig. \ref{bervssnr}, we observe that the BER for uncoded transmission is slightly better than the coded transmission in the lower SNR regime, but as the SNR increases coded transmission performs better than uncoded transmission. For coded transmission, we observe that the proposed allocation schemes (RPE and strip) are performing better than the standard allocation scheme at high SNR. 


\subsection{Optimality of Zak-OTFS Coded Transmission}
\vspace{-0.25em}

    \begin{figure}[htb!]
    \vspace{-0.9em}
        \centering
        \includegraphics[width=0.9\linewidth]{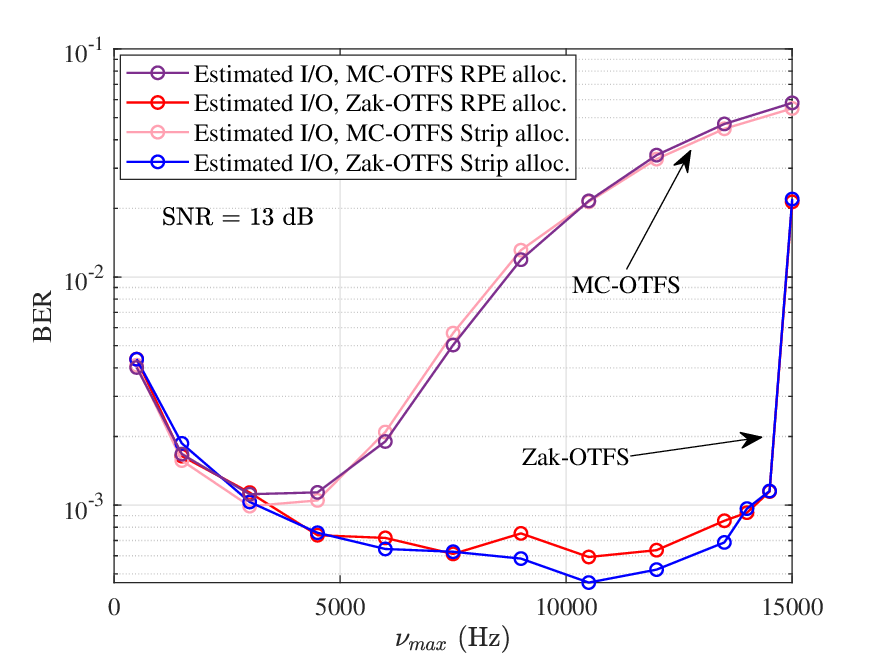}
        \caption{LDPC coding with RPE and strip allocation in MC-OTFS versus Zak-OTFS.}
        \label{mc_vs_zak}
    \end{figure}

Fig. \ref{mc_vs_zak} compares coded BER performance of Zak-OTFS with MC-OTFS with estimated channel, RPE and strip allocation of information symbols, and sinc transmit and receive filters in the DD domain. The maximum $\nu_{max}$ for which reliable communication is possible with MC-OTFS is about $4500$Hz compared with about $14500$Hz for Zak-OTFS. We observe that as $\nu_{max}$ increases the BER for MC-OTFS increases drastically and Zak-OTFS with proposed allocation schemes outperforms.

    \begin{figure}[htb!]
        \centering
        \includegraphics[width=0.9\linewidth]{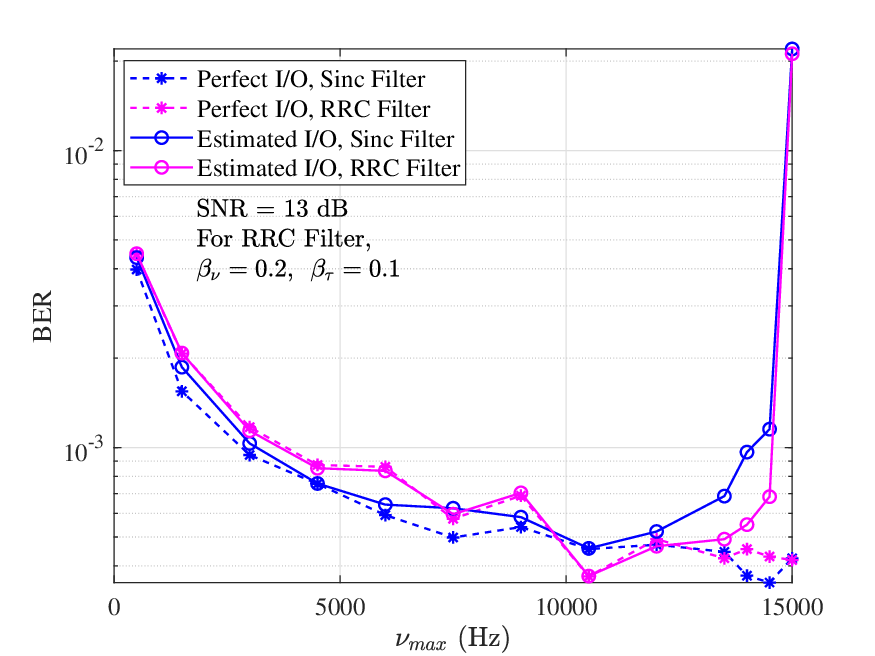}
        \caption{LDPC coding with strip allocation in  Zak-OTFS with sinc versus RRC transmit/receive filters.}
        \label{sincandrrc}
    \end{figure}

Fig. \ref{sincandrrc} compares BER performance of Zak-OTFS with sinc and root raised cosine (RRC) transmit and receive filters using the strip allocation scheme for rate $1/2$ LDPC code. The RRC filter is a factorizable filter specified by a parameter $\beta_\nu$ that controls localization in the Doppler domain and a roll-off parameter $\beta_\tau$ that controls localization in the delay domain \cite{Zakotfs2}. We set $\beta_\tau=0.1$ and $\beta_\nu=0.2$, corresponding to a $10 \%$ increase in bandwidth, and a $20 \%$ increase in data frame duration. We observe that there is little difference in BER performance between the two filters.

\section{Conclusion} \label{sec_conc}
\vspace{-0.3em}
When the crystallization conditions hold, we have explained how acquisition of the Zak-OTFS I/O relation from the response to a single pilot (pulsone) waveform results in channel estimates with different reliabilities, where the precise symbol reliabilities vary with Doppler spread. We have shown that as the Doppler spread increases, the number of highly reliable DD bins decreases, and that the highly reliable DD bins concentrate around the pilot location. We have shown that it is advantageous to allocate information symbols in LDPC codes to more reliable DD bins in the DD domain, and we have proposed a single strip allocation that provides performance gains for all admissible Doppler spreads. By configuring LDPC codes in this way we extend the range of Doppler spreads for which reliable model-free operation is possible on realistic wireless channels. We have also shown that LDPC coding amplifies uncoded performance gains of Zak-OTFS over multicarrier approximations cf. \cite{Zakotfs2}, and that coding makes it possible to use \textit{leaky} transmit and receive filters.

\section{Acknowledgement} 
\vspace{-0.2em}
The work of Robert Calderbank is supported in part by the Air Force Office of Scientific Research under Grants FA 87520-20-2-0504 and FA 9550-20-1-0266, and by the National Science Foundation under Grant FAIN-2148212. Saif Khan Mohammed is also associated with the Bharti school of telecommunication technology and management (BSTTM), IIT Delhi. The work
of Saif Khan Mohammed was supported in part by a project (at BSTTM) sponsored by Bharti Airtel Limited India, and in part by the Prof. Kishan Gupta and Pramila Gupta Chair at IIT Delhi.


\end{document}